\documentclass[a4paper]{jpconf}
\usepackage{geometry}
\usepackage{graphicx}
\usepackage{amssymb}
\usepackage{amsmath}
\usepackage{xcolor}
\usepackage{mathtools}
\usepackage{slashed}
\usepackage{epstopdf}
\usepackage{hyperref}
\usepackage{soul}
\graphicspath{{./}}
\usepackage{ifsym}
\newcommand{\be}{\begin{equation}}                  
\newcommand{\ee}{\end{equation}}                    
\newcommand{\dthreek}{\frac{d^3k}{(2\pi)^3}}        
\newcommand{\Trace}[1]{\mbox{Tr}\left\{#1\right\}}     
\newcommand{\Ima}{\ \mbox{Im}\ }                    
\newcommand{\slsh}[1]{\slashed{#1}}                 
\newcommand{\Gstar}{{}^*G}                          
\newcommand{\Deltastar}{{}^*\Delta}                 

\newcommand{\abs}[1]{|#1|}

\begin{document}
\title{Quark spin - thermal vorticity alignment and the $\Lambda$, $\Bar{\Lambda}$ polarization in heavy-ion collisions}
\author{Alejandro Ayala$^{1,2}$, D. de la Cruz$^{1,3}$,
L. A. Hern\'andez$^{1,2,4}$,\\ S. Hern\'andez-Ort\'iz$^5$ and J. Salinas$^1$}
\address{$^1$ Instituto de Ciencias Nucleares, Universidad Nacional Aut\'onoma de M\'exico, Apartado Postal 70-543, M\'exico, Distrito Federal 04510, Mexico}
\address{$^2$ Centre for Theoretical and Mathematical Physics, and Department of Physics, University of Cape Town, Rondebosch 7700, South Africa}
\address{$^3$ Departamento de F\'isica, Escuela Superior de F\'isica y Matem\'aticas del Instituto Polit\'ecnico Nacional, Unidad Adolfo L\'opez Mateos, Edificio 9, 07738 Ciudad de M\'exico, M\'exico}
\address{$^4$ Facultad de Ciencias de la Educaci\'on, Universidad Aut\'onoma de Tlaxcala, Tlaxcala, 90000, Mexico}
\address{$^5$ Institute of Nuclear Theory,  University of Washington,  Physics-Astronomy Building, Seattle 98195-1550, United States of America}

\ead{ayala@nucleares.unam.mx}
\begin{abstract}
It has been proposed that the $\Lambda$ and $\overline{\Lambda}$ polarizations observed in heavy-ion collisions
are due to the interaction between quark spin and thermal vorticity. In this work we report on a computation of the relaxation time required for this alignment to occur at finite temperature
and baryon chemical potential,  considering quarks with a finite mass. The calculation is performed after modelling the interaction by means of an effective vertex which couples the
thermal gluons and quarks within the vortical medium. We show that the effect of the quark mass is to reduce
the relaxation time as compared to the massless quark case. An intrinsic global polarization of quarks/antiquarks
emerges which is shown to be linked with the $\Lambda$/$\bar{\Lambda}$ polarization.
\end{abstract}
\section{Introduction}
Experiments where heavy-ions collide at relativistic energies provide a unique opportunity to investigate the properties of hadronic
matter under extreme conditions of temperature and density, conditions that are analogous
to those found soon after the Big Bang. At experimental facilities such
as RHIC and the LHC –and in coming years at J-PARC, FAIR, and NICA– two
atomic nuclei collide at relativistic energies giving rise to sequential stages
of strongly interacting matter dynamics, among these, the quark-gluon plasma (QGP) phase. It has been suggested
that during the QGP stage a kinematic vorticity of the order of $\omega \sim 10^{22}$ s$^{-1}$
can be achieved in collisions at RHIC and LHC energies, thus establishing
QGP as the most vortical fluid known so far \cite{STARNature}. The vorticity
is capable to transfer angular momentum to the spin degrees of freedom of
primary quarks, originating the so-called global baryon polarization
once polarized quarks hadronize \cite{LiangWang:2005}. Since QGP can be very
well described using hydrodynamical models, a relativistic hydro description of the
classical vorticity led to the concept of thermal vorticity \cite{Csernai2013,Becattini2013} defined as
\begin{equation}
    \overline{\omega}_{\mu\nu} = \frac{1}{2}\left( \partial_\nu \beta_\mu - \partial_\mu \beta_\nu \right),
\end{equation}
where $\beta_\mu=u_\mu(x)/T(x)$, $u_\mu(x)$ is the local fluid
four-velocity and $T(x)$ is the local temperature. Thermal vorticity describes a rotating fluid that is able to generate a global particle polarization. An outstanding question is what is the time for the spin degrees of freedom to equilibrate with the thermal vorticity. The question is similar to that of finding the time for the momentum degrees of freedom to equilibrate and achieve a common temperature.
\section{Quark Spin - Thermal Vorticity Alignment}
Current hydrodynamical descriptions indicate that the QCD plasma reaches a state of local thermal equilibrium within $\tau_\text{hydro}\sim 1$ fm \cite{UHeinz2015}. After this time, the system can be approximately identified as having a temperature $T$ and quark chemical potential $\mu_q$.
The interaction rate $\Gamma$ of a quark with four-momentum $P=(p_0,\vec{p})$
can be expressed in terms of the quark self energy $\Sigma$ as
\be
    \Gamma(p_0) = \widetilde{f}(p_0-\mu_q)\Trace{\gamma^0\Ima \Sigma},
\ee
where $\widetilde{f}(p_0-\mu_q)$ is the Fermi-Dirac distribution. The
interaction between the thermal vorticity and the quark spin is modeled
by means of a phenomenological effective vertex
\be
    \lambda^\mu_a = g \frac{\sigma^{\alpha\beta}}{2}\overline{\omega}_{\alpha\beta}\gamma^\mu t_a,
\ee
where $\sigma^{\alpha\beta}=\frac{i}{2}[\gamma^\alpha,\gamma^\beta]$ is
the quark spin operator and $t_a$ are the color matrices in the fundamental
representation. 
\begin{figure}[t]
 \begin{center}
  \includegraphics[scale=0.65]{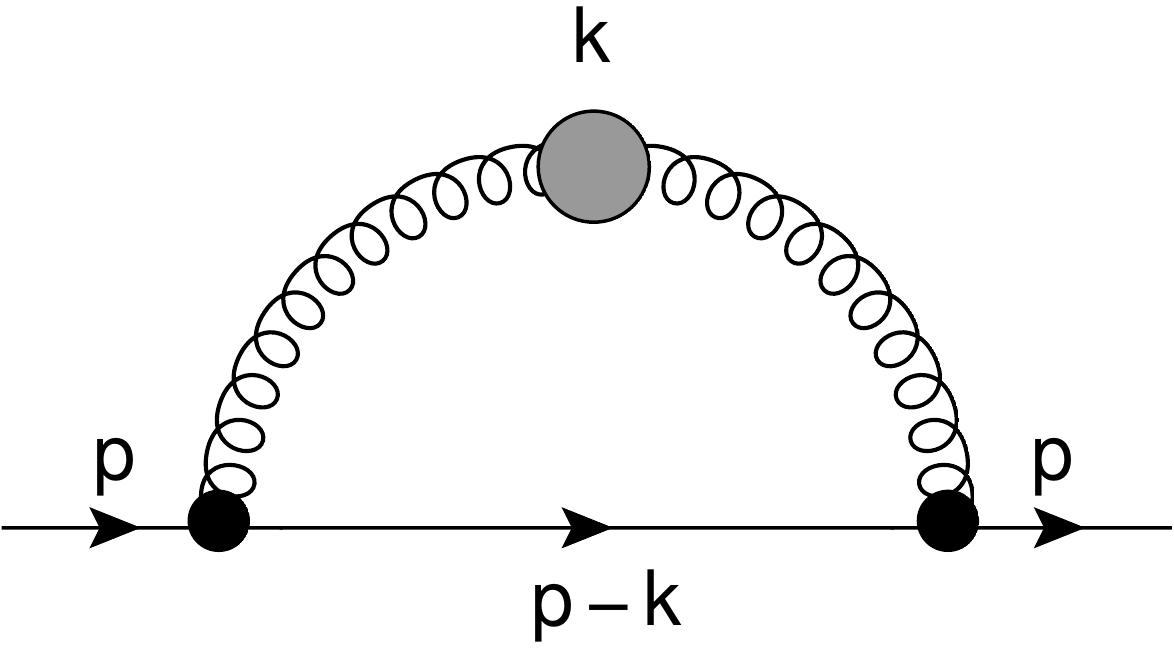}
 \end{center}
 \caption{One-loop Feynman diagram representing the quark self-energy. The gluon line
 with a blob represents the effective gluon propagator at finite density
 and temperature. The blobs on the quark-gluon vertices represent the
 effective coupling between the quark spin and the vorticity. 
 }
 \label{Fig1}
\end{figure}
The one-loop contribution to $\Sigma$, depicted in Fig.~\ref{Fig1},
is given explicitly by
\be
    \Sigma = T\sum_n \int \dthreek\lambda^\mu_a S(\slsh{P}-\slsh{K})\lambda^\nu_b \Gstar^{ab}_{\mu\nu}(K),
\ee
where $S$ and $\Gstar$ are the quark and effective gluon propagators,
respectively.
In a covariant gauge, the Hard Thermal Loop (HTL) approximation to the
effective gluon propagator is given by
\be
   \Gstar_{\mu\nu}(K)=\Deltastar_L(K)P_{L\, \mu\nu} +\Deltastar_T(K)P_{T\, \mu\nu},
\ee
where $P_{L,T\, \mu\nu}$ are the polarization tensors for three dimensional
longitudinal and transverse gluons.
The gluon propagator functions for
longitudinal and transverse modes, $\Deltastar_{L,T}(K)$, are given by
\begin{align}
\Deltastar_L(K)^{-1}&=K^2+2m^2\frac{K^2}{k^2}\left[1-\left(\frac{i\omega_n}{k}\right)Q_0\left(\frac{i\omega_n}{k}\right)\right], \nonumber \\
\Deltastar_T(K)^{-1}&=-K^2-m^2\left(\frac{i\omega_n}{k}\right)\Bigg\{\left[ 1-\left(\frac{i\omega_n}{k}\right)^2\right] Q_0\left(\frac{i\omega_n}{k}\right)+ \left(\frac{i\omega_n}{k}\right)\Bigg\},
\end{align}
where 
\be
Q_0(x)=\frac{1}{2}\ln{\frac{x+1}{x-1}},
\ee
and $m$ is the gluon thermal mass given by
\be
m^2 = \frac{1}{6}g^2C_A T^2 +\frac{1}{12}g^2C_F \left( T^2+\frac{3}{\pi^2}\mu^2\right),
\ee 
where $C_A=3$ and $C_F=4/3$ are the Casimir factors for the adjoint
and the fundamental representations of $SU(3)$, respectively. The sum
over Matsubara frequencies involves products of the propagator functions
for longitudinal and transverse gluons $^*\Delta_{L,T}$ and the Matsubara
propagator for the bare quark $\tilde{\Delta}_F$. Then, the sum over Matsubara frequencies can be expressed as
\begin{eqnarray}
S_{L,T}=T\sum_n\, ^*\Delta_{L,T}(i\omega_n)\tilde{\Delta}_F(i(\omega_m-\omega_n)),
\label{sumprod}
\end{eqnarray}
which is evaluated introducing the spectral
densities $\rho_{L,T}$ and $\widetilde{\rho}$ for the gluon and fermion
propagators, respectively. The imaginary part of $S_i$ (with $i=L,T$) can thus be written as
\begin{equation}
\Ima S_i = \pi\left( e^{(p_0-\mu_q)/T}+1\right)\int_{-\infty}^{\infty}\frac{dk_0}{2\pi}\int_{-\infty}^{\infty}\frac{dp_0'}{2\pi}f(k_0) \widetilde{f}(p_0'-\mu)\delta(p_0-k_0-p_0')\ \rho_i(k_0)\ \widetilde{\rho}(p_0'),
\label{imsum}
\end{equation}
where $f(k_0)$ is the Bose-Einstein distribution. The spectral densities
$\rho_{L,T}(k_0,k)$ are obtained from the imaginary part of $\Deltastar_{L,T}(i\omega_n,k)$
after the analytic continuation $i\omega_n\rightarrow k_0+i\epsilon$ and
contain the discontinuities of the gluon propagator across the real $k_0$-axis. 
Their support depends on the ratio $x=k_0/k$. For $\abs{x}>1$, $\rho_{L,T}$ 
have support on the (time-like) quasiparticle poles. For $\abs{x}<1$ their
support coincides with the branch cut of $Q_0(x)$. On the other hand, the 
spectral density corresponding to a bare quark is given by
\be
\widetilde{\rho}(p_0')=2\pi\epsilon(p_0')\delta(p_0'^2-E_p^2),
\label{kinrest}
\ee 
where $E_p^2=(p-k)^2+m_q^2$ with $m_q$ the quark mass. The kinematical
restriction that Eq.~(\ref{kinrest}) imposes on Eq.~(\ref{imsum}) limits
the integration over gluon energies to the space-like region, namely,
$\abs{x}<1$. Therefore, the parts of the gluon spectral densities that
contribute to the interaction rate are given by
\begin{eqnarray}
\rho_L(k_0,k)&=&\frac{x}{1-x^2}\frac{2\pi m^2\theta(k^2-k_0^2)}{\left[k^2+2m^2\left(1-\frac{x}{2}\ln \left|\frac{1+x}{1-x}\right|\right)\right]^2+\left[\pi m^2x\right]^2},\nonumber \\
\rho_T(k_0,k)&=&\frac{\pi m^2x(1-x^2)\theta(k^2-k_0^2)}{\left\{k^2(1-x^2)+m^2\left[x^2+\frac{x}{2}(1-x^2)\ln \left|\frac{1+x}{1-x}\right|\right]\right\}^2+\left[\frac{\pi m^2x(1-x^2)}{2}\right]^2}.
\end{eqnarray}
Collecting all the ingredients, the interaction rate for a massive quark
with energy $p_0$ to align its spin with the thermal vorticity is given by
\begin{eqnarray}
    \Gamma(p_0) &=& \frac{\alpha_s}{4\pi}\left(\frac{\omega}{T}\right)^2\frac{C_F}{\sqrt{p_0^2-m_q^2}}\int_{0}^{\infty}dk\; k   
    \nonumber \\
    &\quad& \times
    \int_\mathcal{R}dk_0\; [1+f(k_0)]\; \tilde{f}(p_0+k_0-\mu_q)\; \sum_{i=L,T}C_i(p_0,k_0,k)\rho_i(k_0,k),
\label{interactionrate}
\end{eqnarray}
where $\mathcal{R}$ represents the region
\begin{equation}
    \sqrt{\left(\sqrt{p_{0}^{2}
-m_q^{2}}-k\right)^{2}+m_q^{2}}-p_{0}\leq k_{0}\leq \sqrt{\left(\sqrt{p_{0}^{2}
-m_q^{2}}+k\right)^{2}+m_q^{2}}-p_{0}.
\end{equation}
The polarization coefficients $C_{L,T}$
come from the contraction of the polarization tensors $P_{L,T\, \mu\nu}$ with
the trace of the factors involving Dirac gamma matrices from the self-energy. 
After implementing the kinematical restrictions for the allowed values of the
angle between the quark and gluon momenta, these functions are found to be
\begin{align}\label{polarizationcoeffs}
    C_T(p_0,k_0,k) &= 8(p_0+k_0)\, \left(\frac{k^2-2k_0 p_0-k_0^2}{2k\sqrt{p_{0}^{2}-m_q^{2}}}\right)^2, \\
    C_L(p_0,k_0,k) &= -8(p_0+k_0)\left[\left(\frac{k^2-2k_0 p_0-k_0^2}{2k\sqrt{p_{0}^{2}-m_q^{2}}}\right)^2-\frac{1}{2}\right]-8\frac{p_0\, k^2}{k_0^2-k^2}\left(\frac{k^2-2k_0 p_0-k_0^2}{2k\sqrt{p_{0}^{2}-m_q^{2}}}\right)^2.\nonumber 
\end{align}
\begin{figure}[t]
 \begin{center}
  \includegraphics[width=0.65\textwidth]{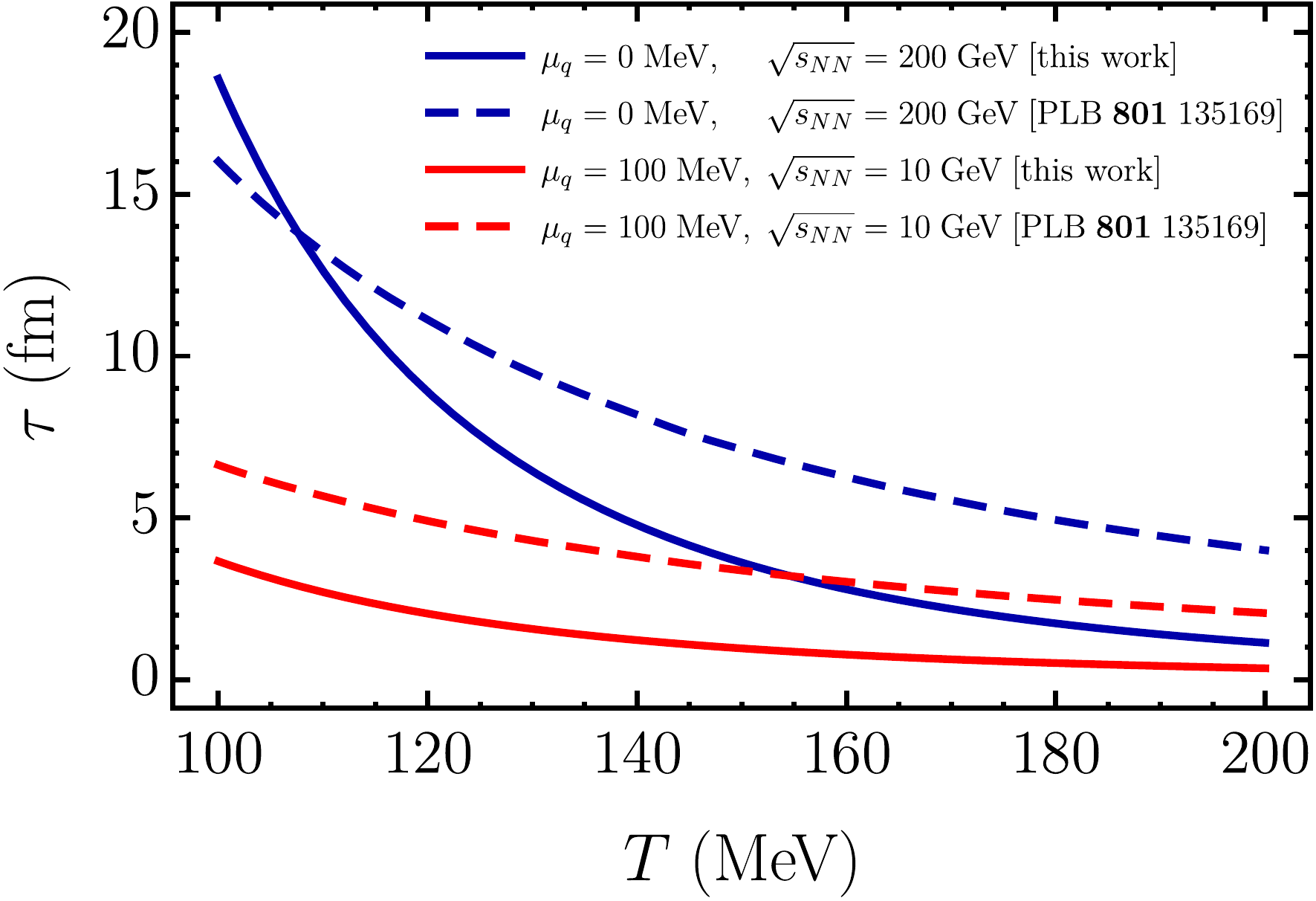}
 \end{center}
 \caption{Relaxation time $\tau$ for quarks as a function of temperature $T$ for semicentral collisions at an impact parameter $b=10$ fm. In dashed lines, massless quarks \cite{Ayala2020} for $\sqrt{s_{NN}}=10,200$ GeV with $\omega\simeq0.12, 0.10$ fm$^{-1}$, respectively. In solid lines, massive quarks for $\sqrt{s_{NN}}=10,200$ GeV with $\omega\simeq0.072, 0.051$ fm$^{-1}$, respectively, using the findings of Ref.~\cite{XDengHuang2020}.}
 \label{Fig2}
\end{figure}
This result should be contrasted with Eqs.~(14) of Ref.~\cite{Ayala2020}, \textit{i.e.},
\begin{align}\label{Cs}
    C_T(p_0,k_0,k) &=8k_0\left(\frac{k^2-2k_0 p_0-k_0^2}{2k p_0}\right)^2, \nonumber\\
    C_L(p_0,k_0,k) &=-8k_0\left[\left(\frac{k^2-2k_0 p_0-k_0^2}{2k p_0}\right)^2-\frac{1}{2}\right],
\end{align}
which were computed for the $m_q\rightarrow 0$ and small quark momentum limit. The
total interaction rate is obtained by integrating Eq.~(\ref{interactionrate})
over the available phase space
\be
\Gamma = V\int \frac{d^3p}{(2\pi)^3}\Gamma(p_0),
\label{reactionrate}
\ee
where $V$ is the volume of the overlap region in the collision. 
Notice that, although the available phase space for a massive quark is reduced as compared to the massless quark case, the
contribution of the new terms in Eqs.~(\ref{polarizationcoeffs}) enhance the overall interaction rate.
Recall that for the collision of symmetric systems of nuclei with radii $R$ and a given
impact parameter $b$, $V$ is given by
\be
   V=\frac{\pi}{3}(4R+b)(R-b/2)^2.
\label{volume}
\ee
\noindent
From the expression for $\Gamma$ in Eq.~(\ref{reactionrate}), we study the parametric dependence of the relaxation time for spin and vorticity alignment, defined as
\be
\tau \equiv 1/\Gamma.
\label{relax}
\ee
\section{Results}
To estimate the angular velocity $\omega$ produced in semicentral collisions, we follow the findings of Ref.~\cite{XDengHuang2020} that provide these values for given energies and impact parameters. Although $\omega$
evolves with time, we work out the computation using its initial value at full nuclei overlap. For an
impact parameter $b=10$ fm, the angular velocity was found to be $\omega \simeq 0.06, 0.04$ fm$^{-1}$
for collision energies $\sqrt{s_{NN}}=10,200$ GeV, respectively.

The relaxation times for quarks of mass $m_q=100$ MeV and quark chemical potential $\mu_q$ 
as a function of temperature are shown in Fig.~\ref{Fig2} for two different energies.
Notice that $\tau\lesssim 5$ fm for the temperature range 150 MeV $<T<200$ MeV, where the phase
transition is expected. In this temperature range, the relaxation times are smaller than
the ones found in Ref.~\cite{Ayala2020} for most of the energy range considered.
Notice that a finite quark mass produces a smaller relaxation time compared to the findings of  Ref.~\cite{Ayala2020}.
\begin{figure}[h]
 \begin{center}
  \includegraphics[width=0.65\textwidth]{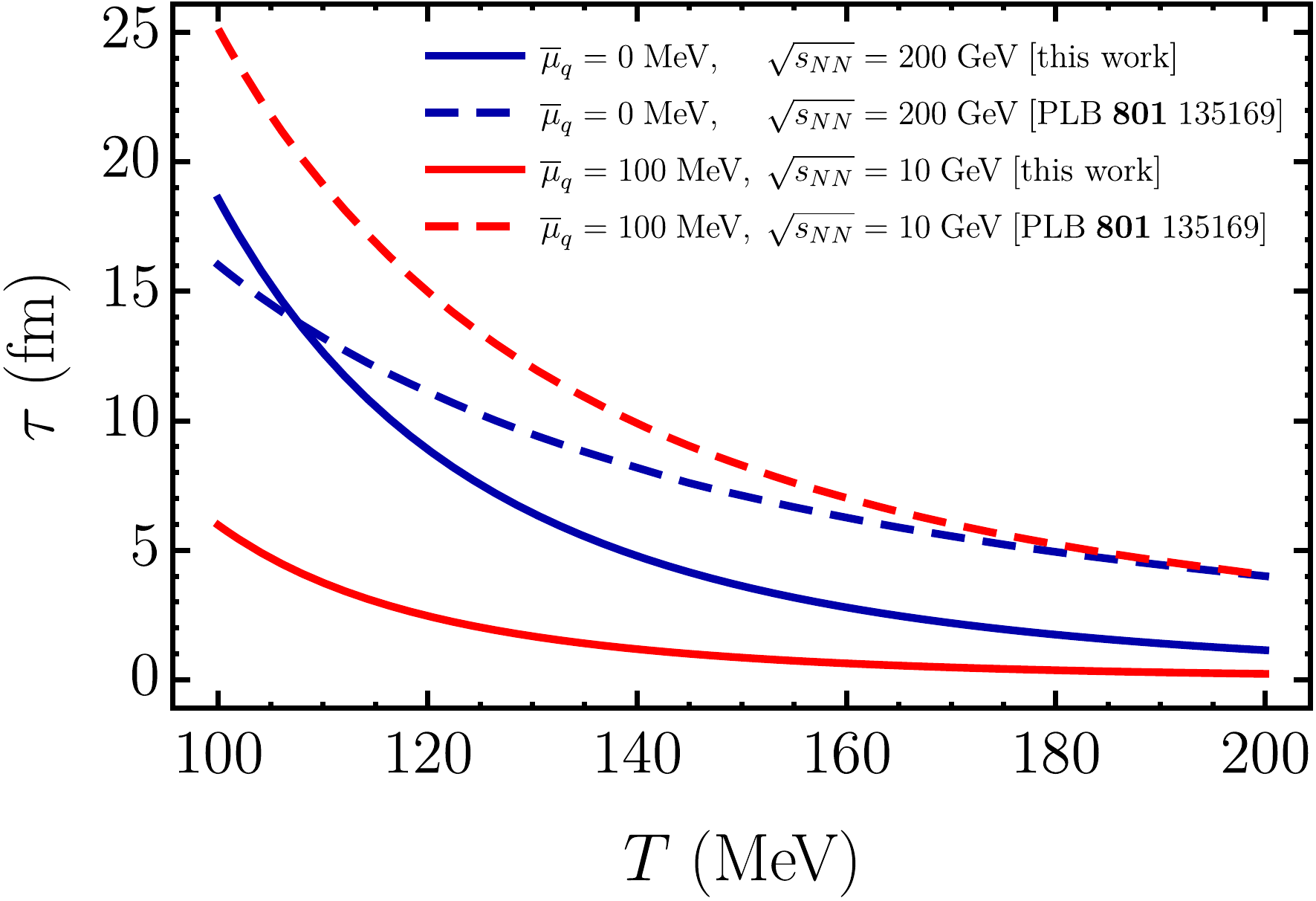}
 \end{center}
 \caption{Relaxation time $\bar{\tau}$ for antiquarks as a function of temperature $T$ for semicentral collisions at an impact parameter $b=10$ fm. In dashed lines, massless quarks \cite{Ayala2020} for $\sqrt{s_{NN}}=10,200$ GeV with $\omega\simeq0.12, 0.10$ fm$^{-1}$, respectively. In solid lines, massive quarks for $\sqrt{s_{NN}}=10,200$ GeV with $\omega\simeq0.072, 0.051$ fm$^{-1}$, respectively, using the findings of Ref.~\cite{XDengHuang2020}.}
 \label{Fig3}
\end{figure}
\par
\begin{figure}[h]
 \begin{center}
  \includegraphics[width=0.5\textwidth]{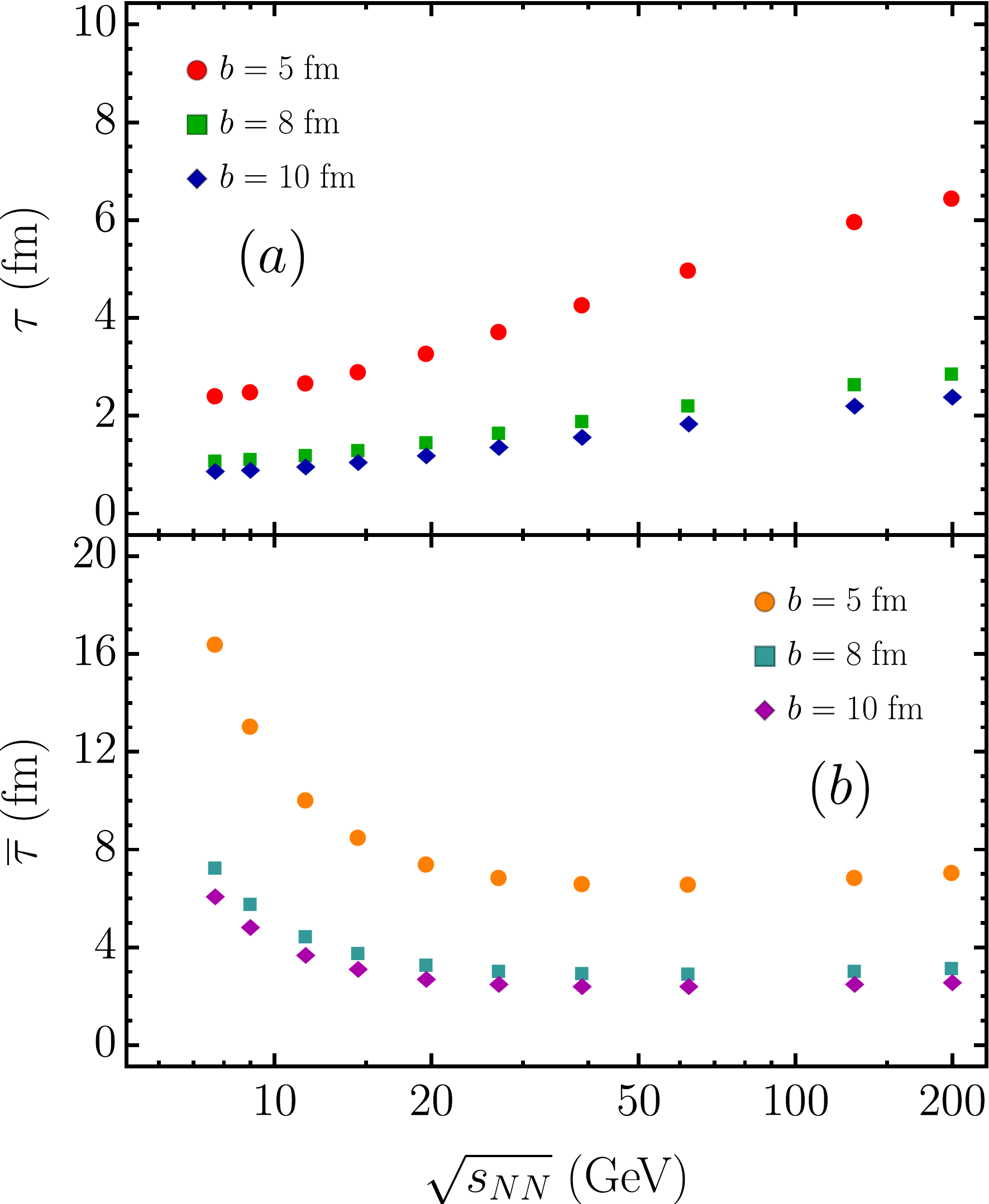}
 \end{center}
 \caption{(a) Relaxation time $\tau$ for quarks as a function of $\sqrt{s_{NN}}$ for semicentral collisions at impact parameters $b=5,8,10$ fm. (b) Relaxation time $\bar{\tau}$ for antiquarks as a function of $\sqrt{s_{NN}}$ for semicentral collisions at impact parameters $b=5,8,10$ fm.}
 \label{Fig4}
\end{figure}

 
For antiquarks with chemical potential $\overline{\mu}_q=-\mu_q$, the resulting relaxation times
for $\sqrt{s_{NN}}=10, 200$ GeV and an impact parameter of $b=10$ fm are shown in Fig.~\ref{Fig3}.
In order to obtain the quark/antiquark relaxation times as a function of collision energy,
we use the freeze-out parametrization of Ref.~\cite{CleymansPRC2006}
\begin{eqnarray}
T(\mu_B)&=&166 - 139\mu_B^2 - 53\mu_B^4,\nonumber\\
\mu_B(\sqrt{s_{NN}})&=&\frac{1308}{1000+0.273\sqrt{s_{NN}}},
\label{cleynm}
\end{eqnarray}
where the freeze-out baryon chemical potential $\mu_B$ and temperature $T$ are given in MeV. These relaxation times are shown in Fig.~\ref{Fig4} Notice that the relaxation times for quarks show a monotonic growth as a function of the collision energy. In contrast, the corresponding relaxation times for antiquarks have a minimum for collision energies in the range 40 GeV $\lesssim\sqrt{s_{NN}}\lesssim$ 70 GeV.
\begin{figure}[t]
 \begin{center}
  \includegraphics[width=0.65\textwidth]{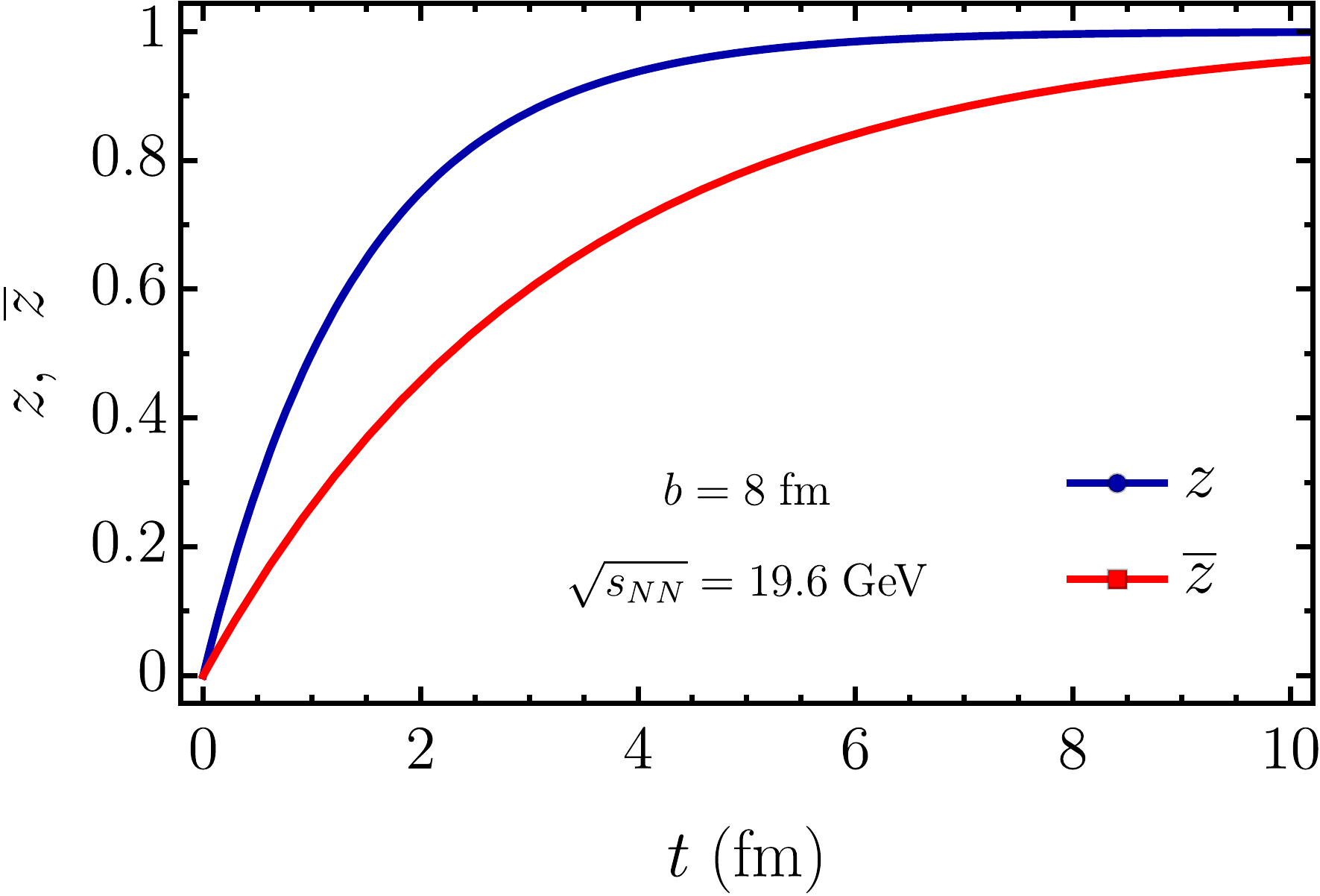}
 \end{center}
 \caption{Intrinsic global polarization for quarks ($z$) and antiquarks ($\bar{z}$) as functions of time $t$ for semicentral collisions at an impact parameter $b=8$ fm for $\sqrt{s_{NN}}=19.6$ GeV.}
 \label{Fig5}
\end{figure}
\par 
Finally, we compute the fraction of globally polarized particles as a function of time, which is given as
\begin{align}
   z&\equiv\frac{N}{N_0}=1-e^{-t/\tau}, \nonumber\\
   \bar{z}&\equiv \frac{\bar{N}}{\bar{N}_0}=1-e^{-t/\overline{\tau}},
\label{intrinsic}
\end{align}
where $z$ and $\bar{z}$ are properly referred to as the {\it intrinsic} global polarization for quarks and antiquarks, respectively. Notice that $\bar{z}<z$ when the impact parameter and the collision energy are the same for both quarks and antiquarks, as can be
seen from Fig.~\ref{Fig5}. Both intrinsic polarizations tend to 1 for $t\simeq 10$ fm. However, a finite intrinsic global polarization for quarks and antiquarks can still be expected when the QGP phase lasts for less than 10 fm.

In conclusion, we have performed a microscopic study to estimate relaxation times for the alignment
between strange quark spin and thermal vorticity at finite temperature and baryon chemical
potential, considering the effects of the quark mass in the alignment. When $T$ and $\mu_B$
are increased and the initial angular velocity is fixed, relaxation times lie well within
the expected life-time of the system, although this does not change the fact that relaxation times
for antiquarks are larger than the corresponding relaxation times for quarks. It is interesting to note that during hadronization, further
effects can enhance $\overline{\Lambda}$ polarization so as to obtain $\mathcal{P}_{\overline{\Lambda}}$
$> \mathcal{P}_\Lambda$. For example, as discussed in this same proceedings~\cite{IvonneWWND} and in more detail in Ref.~\cite{Core-Corona}, this can happen when  the reaction zone is modelled as composed of a high-density core and a less dense corona. The idea was put forward some time ago in Ref.~\cite{Ayala-Cuautle}. Although both regions are subject to the vortical motion, $\Lambda$s and $\overline{\Lambda}$s coming from one or the other regions could show different polarization properties. This can happen since their origins are different: in the core these hyperons come mainly from QGP induced processes. In the corona they come from nucleon-nucleon interaction processes. When this is considered together with a larger abundance of $\Lambda$s as compared to $\overline{\Lambda}$s in the corona and a smaller number of $\Lambda$s coming from the core as compared to those coming from the corona, an amplification effect for the $\overline{\Lambda}$ polarization can occur. This is more prominent for semi-central to peripheral collisions and small collision energies. Further details are provided in Refs.~\cite{IvonneWWND, Core-Corona}.
\ack Support for this work has been received in part by UNAM-DGAPA-PAPIIT grant number IG100219 and by Consejo Nacional de Ciencia y Tecnolog\'ia grant numbers A1-S-7655 and A1‐S‐16215.

\end{document}